\documentclass{article}
\usepackage{amsmath}                             
\usepackage{amsfonts}
\usepackage{epsfig}
\input psfig.sty
\def\fig{.}

\def\n{\noindent}

\begin{document}

\baselineskip .7cm

\author{Navin Khaneja \thanks{To whom correspondence may be addressed. Email:navin@hrl.harvard.edu} \thanks{Division of Engineering and Applied Sciences, Harvard University, Cambridge, MA 02138. This work was funded by DARPA QUIST grant 496020-01-1-0556, NSF 0218411, NSF 0133673.} ,\ \ Jr-Shin Li$\ ^\dagger$, \ Cindie Kehlet$\ ^\ddagger$,\ \ Burkhard Luy$\ ^\ddagger$,\ \ Steffen J.
Glaser
\thanks{Department of Chemistry, Technische Universit\"at M\"unchen,
85747 Garching, Germany.  This work was funded by the Fonds der 
Chemischen Industrie and the Deutsche Forschungsgemeinschaft under 
grant Gl 203/4-2.}}

\title{\bf Broadband Relaxation-Optimized Polarization Transfer in Magnetic Resonance}

\maketitle

\vskip 1cm

\begin{center}
{\bf Abstract}
\end{center}
\n 
Many applications of magnetic resonance are limited by rapid loss of spin coherence 
caused by large transverse relaxation rates. In  nuclear magnetic
resonance (NMR) of large proteins, increased relaxation losses lead
to poor sensitivity of experiments and increased measurement time. In this paper 
we develop broadband relaxation optimized pulse sequences (BB-CROP) which approach
fundamental limits of coherence transfer efficiency in the presence of very general relaxation mechanisms
that include cross-correlated relaxation. These broadband transfer schemes use new techniques of chemical shift
refocusing (STAR echoes) that are tailored to specific trajectories  of coupled spin evolution. 
We present simulations and experimental data indicating significant enhancement in the
sensitivity of multi-dimensional NMR experiments of large molecules by use of these methods.

\vskip 3em

\newpage
\section{Introduction}

The loss of signal due to spin relaxation \cite{Redfield} is a major problem in many
practical applications of magnetic resonance. An important application is NMR spectroscopy 
of proteins \cite{wuthrichbook, palmer}. Multidimensional coherence transfer experiments in
protein NMR are characterized by large transverse relaxation rates. When these 
relaxation rates become comparable to the spin-spin couplings, the efficiency of coherence transfer is considerably
reduced, leading to poor sensitivity and  limiting the size of macro molecules that can be
analyzed by NMR. Recent advances have made it possible to significantly extend the size limit of
biological macro molecules amenable to study by liquid state NMR 
%\cite {TROSY1, TROSY2, TROSY3, CRINEPT}
[4-7].
These techniques take advantage of the phenomenon of cross-correlation or interference between
two different relaxation mechanisms 
%\cite{CC1, CC2, CC3, CC4, CC5, CC6}
[8-13]
Until recently, it was not clear if further improvements can
be made and what is the physical limit for the coherence transfer efficiency between coupled spins in
the presence of cross-correlated relaxation. In our recent work, using methods from optimal
control theory, we derived fundamental limits on the efficiency of polarization transfer in the
presence of general relaxation mechanisms 
%\cite{Crop, Rope, zRope}
[14-16]. This established that
state of the art experiments in NMR have the potential for significant improvement. We also
provided relaxation-optimized pulse sequences which achieve the theoretical maximum transfer
efficiency for a single spin pair. However, in order to apply these methods to practical NMR
experiments, one needs to simultaneously address a family of 
 coupled spin pairs with
dispersion in their Larmor frequencies.
In the limiting cases where
cross-correlation rates are either much smaller or much larger than the spin-spin coupling, 
modifying the narrow-band relaxation optimized pulses into broadband transfer schemes is straight-forward 
by use of conventional
refocusing techniques. However, 
in experiments, 
where both coupling and cross-correlation rates are
comparable, the use of
conventional refocusing methods for making relaxation optimized sequences broadband 
significantly reduces the transfer efficiencies as these methods eliminate either the 
spin-spin couplings or the cross-correlation effects.
Finding broadband transfer schemes which can achieve the efficiency of relaxation-optimized
sequences required the development of  \underline{s}pecific
\underline{t}rajectory \underline{a}dapted \underline{r}efocusing (STAR) methods, where refocusing is performed
in a moving coordinate system attached to an optimal  trajectory.
In this paper, we present these new methods and resulting
broadband relaxation-optimized polarization transfer experiments.

\section{Theory}

We consider an isolated heteronuclear spin system
consisting of two coupled spins 1/2, denoted  $I$ (e.g. $^1$H) and $S$ (e.g. $^{15}$N). 
We address the problem of 
selective population inversion of two
energy levels (e.g. $\alpha \beta$ and $\beta \beta$) as shown in Fig. 1. 
This is a central step in high-resolution multi-dimensional NMR spectroscopy \cite{Ernst} and
corresponds to the transfer of an initial density operator $I_z$, representing polarization on spin
$I$, to the target state $2I_z S_z$, representing two-spin order.
\begin{center}
\begin{figure}[h] 
\centerline{\psfig{file= \fig/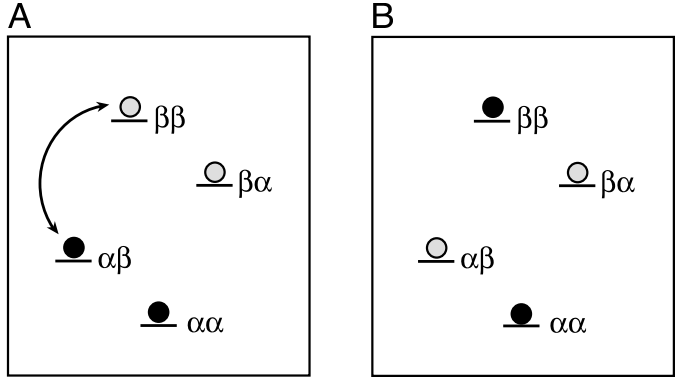 ,width=3.5in}}
\caption{\label{fig:beta_etc} 
The broadband transfer of polarization
$I_z$ (A) to
$ 2I_z S_z$ (B) corresponds to an offset-independent, but transition-selective population inversion
of the energy levels $\alpha \beta$ and $\beta \beta$} 
\end{figure}
\end{center}
For large molecules in the so-called spin diffusion limit \cite{Ernst}, where longitudinal
relaxation rates are negligible compared to transverse relaxation rates, both
the initial term ($I_z$) and final term ($2 I_z S_z$) of the density operator are long-lived.
However, the transfer between these two states requires the creation of coherences which in general
are subject to transverse relaxation.
The two principal transverse relaxation mechanisms are dipole-dipole (DD)
relaxation and relaxation due to chemical shift anisotropy (CSA) of spins $I$ and $S$.
The quantum mechanical equation of motion (Liouville-von Neumann equation) for the
density operator $\rho$ \cite{Ernst} is given by
\begin{eqnarray}\label{eq:rho_dot} 
\nonumber
\dot{\rho} &=& \pi \ J [-i 2  I_z S_z, \rho] 
+ \pi \ k_{DD} [2 I_z S_z, [2 I_z S_z , \rho]]
+ \pi \ k^{I}_{CSA} [I_z, [I_z , \rho]]
 + \pi \ k^S_{CSA} [S_z, [S_z , \rho]] \\
 &\ &
+ \ \pi \ k^{I}_{DD/CSA} [2 I_z S_z, [I_z , \rho]] 
 + \pi \ k^{S}_{DD/CSA}  [2 I_z S_z, [S_z , \rho]],
\end{eqnarray}
where $J$ is the heteronuclear coupling constant.
The rates $k_{DD}$, $ k^{I}_{CSA} $, $ k^{S}_{CSA}$ represent auto-relaxation rates due to DD
relaxation, CSA relaxation of spin $I$ and CSA relaxation of spin $S$, respectively. The rates
$k^{I}_{DD/CSA}$ and
$k^{S}_{DD/CSA}$ represent
 cross-correlation rates of spin $I$ and $S$ caused by interference effects between DD and CSA
relaxation. These relaxation
rates
depend on various physical parameters, such as the gyromagnetic ratios of the spins, the
internuclear distance,  the CSA tensors, the strength of the magnetic field and the correlation time
of the molecular tumbling
\cite{Ernst}. Let the initial density operator $\rho(0) = A$ and $\rho(t)$ denote the density operator at time $t$. The maximum efficiency of 
transfer between $A$ and a target operator $C$ is defined as the largest possible value of
Trace$\{C^{\dagger} \rho(t)\}$ for any time $t$ \cite{Science} (by convention operators A and
C are normalized).

\begin{center}
\begin{figure}[h]
\centerline{\psfig{file= \fig/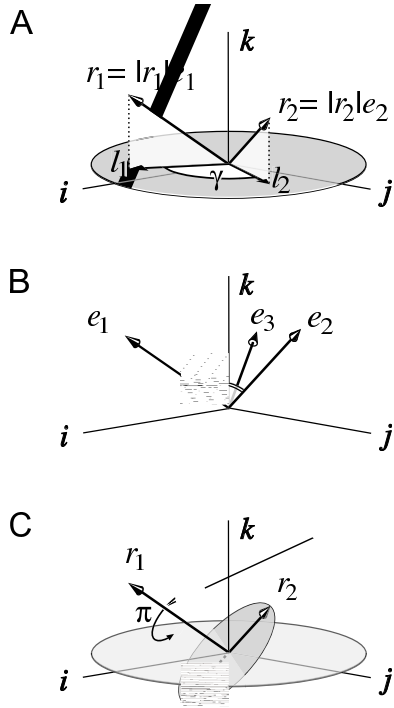 ,width=2.0in}}
\caption{\label{fig:2}
(A) Schematic representation of the magnetization vector $r_1=(\langle I_x \rangle,
 \langle I_y\rangle, \langle I_z\rangle)$ and of the antiphase vector  $r_2=(\langle 2 I_x
S_z\rangle,
 \langle 2 I_y S_z\rangle, \langle 2 I_z S_z\rangle)$ in the common frame spanned by the
standard Cartesian unit vectors ${\bf i}$, ${\bf j}$, and ${\bf k}$. The vectors $l_1$ and $l_2$
are the projections of $r_1$ and $r_2$ into the transverse plane and $\gamma$ is the angle between
$l_1$ and $l_2$. (B) For the optimal CROP (cross-correlated relaxation optimized pulse) trajectory
\cite{Crop}, the units vectors
$e_1$,
$e_2$  in the direction of $r_1$ and $r_2$ are orthogonal and together with
$e_3 = e_1 \times e_2$  define a specific moving frame along the optimal trajectory. 
(C) The pulse element
 $R_1(t)$ consists of a
$\pi$ rotation of spin $I$  around $e_1$, which leaves $r_1$ invariant and inverts $r_2$
(dashed arrow) and a $\pi$ rotation of spin $S$ around an arbitrary axis in the transverse plane,
which also leaves $r_1$ invariant and brings $r_2$ back to its initial position (solid arrow).
Hence, $R_1$ neither changes the ratio $\vert l_2 \vert / \vert l_1\vert $ nor the angle
$\gamma$ which are both constants of motion for the optimal CROP trajectory.} 
\end{figure}
\end{center}

In our recent work \cite{Crop} we showed that for a single spin pair $IS$, the maximum
efficiency
$\eta$ of transfer between the operators $I_z$ and $2I_z S_z$ depends only on the scalar coupling
constant
$J$ and the net auto-correlated and cross-correlated relaxation rates of spin $I$, given by 
$ 
k_a=k_{DD}+ k^{I}_{CSA}  
$
and  
$
k_c=k^{I}_{DD/CSA}
$, respectively.
Here the rates $k_a$ and $k_c$ are a factor of $\pi$ smaller than in conventional definitions of
the rates, e.g., $k_a=1/(\pi T_2$), where $T_2$ is the transverse relaxation time in the absence of
cross-correlation effects \cite{Crop, Rope}. The physical limit
$\eta$ of the transfer efficiency is given by
\cite{Crop}
\begin{equation}\label{eq:efficiency} 
\eta = 
\sqrt{1+\zeta^2} - \zeta, \end{equation}  
where
$
\zeta^2= {({k_a^2 - k_c^2})/({J^2 + k_c^2})}.
$
The optimal transfer scheme (CROP: cross-correlated relaxation optimized pulse)
has two constants of motion (see Figure 2 A).  If $l_1(t)$ and $l_2(t)$ denote the two-dimensional
vectors
$(\langle I_x \rangle(t),
 \langle I_y\rangle(t))$ and $(\langle 2 I_x S_z \rangle(t),
 \langle 2 I_y S_z \rangle(t))$, respectively, then 
throughout the optimal
transfer process the ratio $\vert l_2\vert / \vert l_1  \vert$ of the magnitudes of the vectors
$l_2$ and
$l_1$ should be maintained constant at
$\eta$. Furthermore, the angle $\gamma$ between $l_1$ and $l_2$ is constant throughout.
These two constants of motion depend on the transverse relaxation rates and the coupling constants
and can be explicitly computed \cite{Crop}.
These constants  determine the amplitude and phase
of the rf field at each point in time and explicit expressions for the optimal pulse sequence can be
derived.
In Fig. 3 A and B, the optimal rf amplitude and phase
 of a CROP sequence is shown as a function of time  for the case $k_c/k_a=0.75$ and
$k_a=J$. 

\begin{center}
\begin{figure}[h] 
\centerline{\psfig{file= \fig/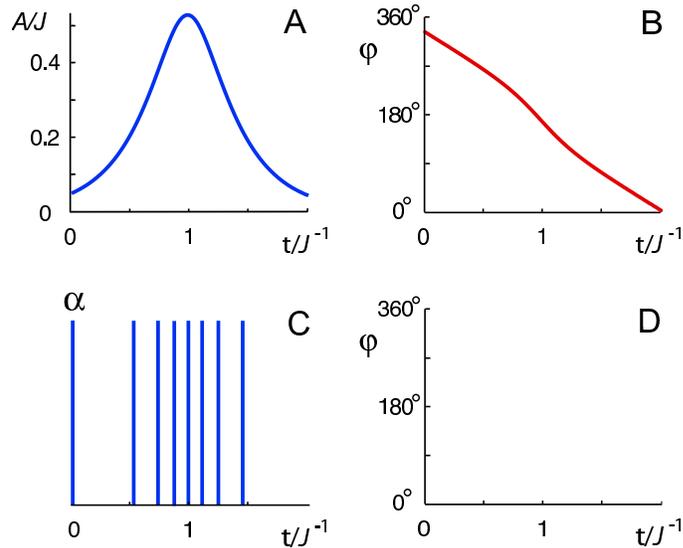 ,width=3.5in}}
\caption{\label{fig:beta_etc}
 Ideal (A, B) and approximate (C, D) implementations of 
an on-resonance CROP sequence \cite{Crop} for 
$k_a=J$ and $k_c/k_a=0.75$.
Panel A shows the ideal rf amplitude $A(t)=-\gamma_I B^I_1(t)/(2 \pi)$ (where $\gamma_I$ is
the gyromagnetic ratio of spins $I$) in units of the coupling constant
$J$ and panel C shows a schematic representation of an approximate CROP sequence consisting of 8
hard pulses of flip angle $\alpha= 21.5^\circ$. Panels B and D show the phases 
$\varphi(t)$ of the ideal CROP sequence and its hard pulse approximation.
}
\end{figure}
\end{center}

The transfer scheme as described assumes that the resonance frequencies of a
single spin pair are known exactly. 
Therefore, the above methods 
cannot be directly used in spectroscopic applications with many spin pairs and a dispersion of
Larmor frequencies. In this paper, we  develop methods to make the above principle of relaxation
optimized  transfer applicable for a broad frequency range, making these 
methods suitable for spectroscopy of large proteins. A straightforward method of converting the 
smooth pulse shapes (like Fig. 3A) into a broadband transfer scheme can be realized by the following steps.

\n a) Given the optimal amplitude $A(t)$ and phase $\varphi(t)$, of the on-resonance 
pulse (see Fig. 3 A and 3 B), we can approximate the smooth pulse shape as a sequence of hard pulses 
with small flip angles $\alpha_k$ separated by evolution periods of duration $\Delta_k$  (c.f. Figs. 3 C). 
These are DANTE-type sequences (delays alternating with nutations for tailored
excitation) \cite{Morris:Dante}. The flip  angle $\alpha_k$ at time $t$ is just  
$ \int_{t}^{t +  \Delta_k} A(\tau)  d\tau, $ with the phase given by 
$\varphi_k =\varphi(t)$ (c.f. Fig. 3 D). The delays $\Delta_k$ could be
chosen in many ways. For example, they may be all equal or can be chosen so that the 
flip angles $\alpha_k$ are equal (c.f. Figs. 3 C).
 
\n b) Insertion of $\pi$ pulses in the center 
of delays to refocus the transverse components of the spins \cite{zzz-Paper}, see Fig. 4 A-C.

\begin{center}
\begin{figure}[h] 
\centerline{\psfig{file= \fig/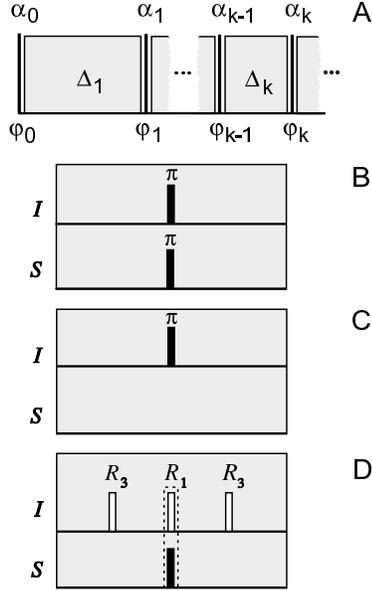 ,width=1.9in}}
\caption{\label{fig:beta_etc2} 
(A) Architecture of broadband relaxation-optimized pulse sequences, consisting of $N$ periods
of duration $\Delta_k$ (gray boxes, shown in panes B-D in more detail) and
 hard rf pulses
with flip angles $\alpha_k$ and phases $\varphi_k$. (B) Chemical shift refocusing scheme
preserving transfer through $J$ coupling but eliminating transfer though the  cross-correlated
relaxation rate $k_c$. (C) Chemical shift refocusing scheme
preserving transfer through $k_c$  but eliminating transfer though $J$.
(D) STAR echo scheme
preserving both transfer through $k_c$  and though $J$. In B, C, and D, black bars represent
180$^\circ$ rotations around an axis in the x-y plane, white bars represent 180$^\circ$ rotations
around tilted axes.
 }
\end{figure}
\end{center}

Note that this method of making relaxation optimized pulses broadband is only applicable if one
is using either just the couplings (as in INEPT \cite{INEPT2} 
or ROPE \cite{Rope} transfer) or just the cross-correlation effects (as in standard CRIPT \cite{cript3} 
or CROP \cite{Crop} transfer for $J=0$) as the transfer mechanism.

For example, the relaxation-optimized pulse elements (ROPE) \cite{Rope}, which only use transfer
through couplings (special case of CROP \cite{Crop} when $k_c=0$) can be made broadband in a 
straight-forward way as explained above. Simultaneous
$\pi$ rotations applied to spins $I$ and $S$ in the middle of the evolution periods refocus
the chemical shift evolution while retaining the coupling terms (see Fig. 4 B). Note however,
that such a pair of $\pi$ rotations will eliminate any DD-CSA cross-correlation effects that might be
present
\cite{CRINEPT}.

On the other hand,  if
$J$ is very small or
$k_c$ is close to $ k_a$ (in which case transfer using cross-correlation effects is
very efficient, c.f. Eq. \ref{eq:efficiency}), it is desirable to
use relaxation-optimized sequences which
only use cross-correlation effects for transfer (special case of CROP \cite{Crop} when $J=0$).
Such a relaxation optimized transfer is characterized by a smooth rotation $I_z \rightarrow I_x$
and vice versa ($-2I_xS_z \rightarrow
2I_zS_z$). Again such a transfer can be made broadband as explained above. In this case 
the refocusing $\pi$ pulses are applied only to spin $I$ in the center of delays (see Fig. 4 C).
By such pulses, cross-correlation effects are retained but coupling evolution is eliminated \cite{CRINEPT}.
%(c.f. Appendix ...). 

Therefore
the advantage of the CROP pulse  sequence (which simultaneously uses both $J$ couplings and
cross-correlation effects) would be lost in using this conventional strategy to make these
sequences broadband.   
The key observation for making CROP transfer
broadband is that in the on-resonance CROP transfer scheme, the 
magnetization vector $$ r_1(t) = \langle I_x\rangle(t)\ {\bf i} + \langle 
I_y\rangle(t)\ {\bf j} + \langle I_z\rangle(t)\ {\bf k} $$ always remains 
perpendicular (c.f. Fig. 2 A and B) to the net antiphase vector 
$$ r_2(t) = \langle 2I_x S_z\rangle(t) \ {\bf i} + \langle 2I_y 
S_z\rangle(t) \ {\bf j} + \langle 2I_z S_z\rangle(t)\ {\bf k}, $$ 
where ${\bf i}$, ${\bf j}$, and ${\bf k}$ are the standard Cartesian unit vectors
(for  details see {\it Supporting Methods}).  Let $e_1$, $e_2$ denote unit 
vectors in the direction of $r_1$ and $r_2$ and let $e_3 = e_1 \times e_2$ 
denote the unit normal pointing out of the plane spanned by $e_1$ and $e_2$. 

%Let $R_I(e, 
%\theta)$ denote counterclockwise rotation of spin $I$ around the direction 
%$e$ by an angle $\theta$. 
%Observe the rotation 
%$R_I(e_1(t), \pi)$ leaves $e_1(t)$ intact while $r_2(t) \rightarrow 
%-r_2(t)$ . A $\pi$ rotation of spin $S$ around the x axis, denoted by $R_S(x, \pi)$ has the 
%same effect of  $r_2(t) \rightarrow -r_2(t)$. We concatenate the two 
%rotations in 

\noindent
Let $R_1(t)$ denote a $\pi$ rotation of spin $I$  around $e_1(t)$ and a simultaneous 
$\pi$ rotation of spin $S$ around an arbitrary axis in the transverse plane. Observe that 
$R_1(t)$ fixes the vectors $r_1(t)$ and
$r_2(t)$, see Fig. 2 C. 
Similarly, 
let $R_2(t)$ denote a $\pi$ rotation around $e_2(t)$ and a simultaneous 
$\pi$ rotation of spin $S$ around an arbitrary axis in the transverse plane.
$R_2(t)$ inverts $r_1(t)$ 
and $r_2(t)$, i.e. $r_1 (t) \rightarrow -r_1(t)$ and $r_2(t) \rightarrow 
-r_2(t)$.
We also define   
$R_3(t)$ as a $\pi$ rotation around $e_3(t)$  which also results in $r_1 (t)
\rightarrow -r_1(t)$ and $r_2(t) 
\rightarrow -r_2(t)$.
Note that these rotations are special because they neither change the ratio
$\vert l_2\vert / \vert l_1\vert $ nor the angle $\gamma$ between the transverse components $l_1$
and
$l_2$.

We now show how the rotations $R_1$ and $R_3$ can be used to produce a 
broadband cross-correlated relaxation optimized  pulse (BB-CROP) sequence. Given the implementation of the on resonance CROP pulse
(Fig. 3 A and 3 B) as a sequence of pulses and delays (Fig. 3 C and 3 D), the chemical shift evolution during a delay $\Delta$ can be refocused by the
sequence (c.f. Fig. 4 D)
$$ {\Delta \over 4}  R_3 {\Delta \over 4} R_1 {\Delta \over 4} R_3 {\Delta \over 4}. $$
The rotations $R_1(t)$ and $R_3(t)$ are defined using the optimal trajectory and
keep changing from one delay to another, as the vectors $r_1(t)$ and
$r_2(t)$ evolve.
We refer to this specific trajectory adapted refocusing as STAR. To analyze how this refocusing works,
at time instant $t$ consider the coordinate 
system defined by $e_1(t)$, $e_2(t)$ and $e_3(t)$ (c.f. Fig. 2 B). The unit vector along 
$z$ can be written as $ae_1(t) + be_2(t) + ce_3(t)$.
The chemical shift 
evolution generator $I_z$ can be 
expressed as  
\begin{equation}\label{eq:izdecomp} 
I_z=a I_{e_1} + b I_{e_2} + c I_{e_3}
\end{equation}  
and the evolution for time $ { {\Delta \over 4}}$ 
under the chemical shift takes the form $\exp \{- {\rm i}\ \omega (a I_{e_1} + b 
I_{e_2} + c I_{e_3})  {\Delta \over 4}\}$. 
Assuming that the $R_3$ rotation is fast, so that there is negligible chemical 
shift evolution (and negligible relaxation) during the $R_3$, the sequence $ {\Delta \over 4} \ R_3
\ 
 {\Delta \over 4} $ produces the net evolution 
$$\exp \{- {\rm i}\ \omega (a I_{e_1} + b I_{e_2} + c I_{e_3})  {\Delta \over 4}\} \ \ R_3 \  \exp
\{- {\rm i}\
\omega  (a I_{e_1} + b I_{e_2} + c I_{e_3})  {{\Delta} \over {4}}\} \ \hskip 2em $$
$$  \ \hskip 2em =R_3 \exp \{- {\rm i}\ \omega (- a I_{e_1} - b I_{e_2} + c I_{e_3})  {\Delta
\over 4}\}
\ 
\exp
\{- {\rm i}\
\omega  (a I_{e_1} + b I_{e_2} + c I_{e_3})  {\Delta \over 4}\}.$$
For delays $\Delta\ll 1/\omega$, the effective evolution can be approximated by $R_3  \exp
\{- {\rm i}\
\omega
 \ c\   I_{e_3} {\Delta \over 2}\}$. Now the rotation $R_1$ %or $R_2$ 
can be used to refocus the remaining chemical shift evolution 
due to $I_{e_3}$ by the complete STAR echo sequence
${\Delta \over 4} R_3 {\Delta \over 4} R_1 {\Delta \over 4} R_3 {\Delta \over 4}$. The effective
evolution during the period $\Delta$  $$  R_1 \exp \{ {\rm i}\ \omega c I_{e_3}{\Delta \over
2}\}\exp
\{-{\rm i}\
\omega c I_{e_3}{\Delta \over 2}\}
\approx R_1, 
$$ 
i.e. chemical shift evolution is eliminated. Note, we assume that the frame $e_1$,
$e_2$, $e_3$ does not evolve much during the four
${\Delta
\over 4}$ periods so that the two $R_3$ rotations are approximately the same.
Under this STAR sequence, the general coupling evolution exp$\{-i 2 \pi J I_z S_z\}$  and the
general Liouvillian evolution (containing cross correlation effects) is not completely preserved. 
Inspite of this, the evolution of $r_1(t)$ and $r_2(t)$ for the CROP trajectory is unaltered. 
This is because, for this specific trajectory, the magnitude of the transverse components $l_1(t)$ and $l_2(t)$
and the angle $\gamma$ between them is not changed by application of these tailored refocusing pulses.
Since all evolution is confined to transverse operators, the efficiency of the BB-CROP pulse is unaltered 
by application of STAR refocusing pulses.

\section{Practical considerations}

180$^\circ$ rotations around tilted axes as required by the operations ($R_1$ and $R_3$)
of the STAR echo method can be realized in practice by off-resonance pulses. For example,  a
180$^\circ$ rotation around an axis forming an angle $\theta$ with the $x$
axis can be implemented by a pulse with an rf amplitude $\nu_1$ and offset $\nu_{\rm
off}=\nu_1\tan
\theta$ with a pulse duration $\tau_p=1/(2 \nu_{\rm eff})$, where $\nu^2_{\rm eff}=\nu_1^2 +
\nu^2_{\rm off}$. 
At the start of the pulse, we assume that both the on-resonance and off-resonance rotating frames
are aligned. In the off-resonance rotating frame the axis of rotation does not move.
After the pulse, the off-resonant rotating frame has acquired an angle of 
$\phi_{\rm off} =2 \pi \nu_{\rm off} \tau$ 
 relative to the on-resonance frame. For a pulse
sequence specified in the on-resonance rotating frame, this can be taken into account by adding
the phase
$\phi_{\rm off}$ acquired during a given off-resonant 180$^\circ$ pulse to the nominal phases of all
following pulses on the same rf channel (In the sequence provided in supporting methods, this correction 
has been incorporated). Alternative implementations of rotations around tilted
axes by composite on-resonance pulses would be longer and could result in larger relaxation losses
during the pulses.

\n Under the assumption of ideal impulsive 180$^\circ$  rotations (with negligible pulse duration
and negligible rf inhomogeneity), the STAR approach realizes a broadband transfer of
polarization that achieves the optimal efficiency as given in
\cite{Crop}. However,  spectrometers are limited in terms of their maximum rf amplitude and
homogeneity of the rf field. Therefore in practice, pulses have finite widths and hence evolution
(especially relaxation) becomes important during the pulse duration. The effect becomes
pronounced as the number of 180$^\circ$ pulses is increased in order to keep the
refocusing periods
$\Delta_k$ short for a better approximation to the on-resonance CROP pulse.
We observe that after a point the loss caused due to relaxation during pulse periods
overshadows the gain in efficiency one would expect by finer and finer approximations of the
ideal CROP trajectory.
Furthermore, dephasing due to rf inhomogeneity increases as the number of 180$^\circ$  pulses
is increased. Therefore one is forced to find a compromise between loss due to a large
number of 180$^\circ$ pulses versus: (a) loss of efficiency due to a coarser  discretization of
the CROP pulse, (b) reduced bandwidth of frequencies that can be refocused by an
increased duration of the refocusing periods.
When the number of refocusing periods becomes small, it is important to find a good way to
discretize the CROP pulse so as to maximize the efficiency of coherence transfer that can be
achieved by a pulse sequence with a prescribed number of evolution periods. We have developed
rigorous control theoretic methods  based on
the principle of dynamic programming \cite{Bellman} to efficiently achieve this discretization (see {\it Supporting Methods}). 
This helps us to compute optimal approximations of  CROP pulse sequences as a series of a small number of 
pulses and delays very efficiently.

\begin{center}
\begin{figure}[h] 
\centerline{\psfig{file= \fig/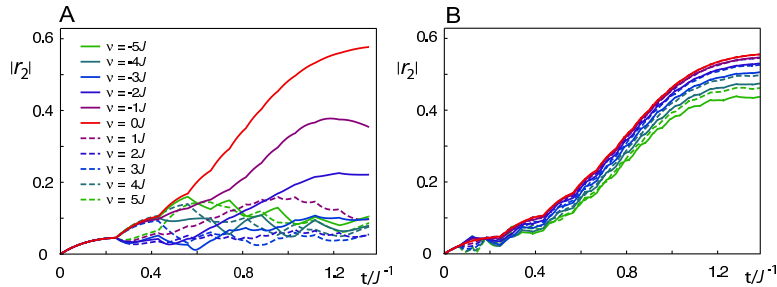 ,width=4.0in}}
\caption{\label{fig:beta_etc5} 
The
buildup of antiphase vectors $r_2$  is shown for 11
different offset frequencies in the range of $\pm 5 J$
during (A) a selective CROP (without STAR echoes) and
(B) a corresponding BB-CROP (with STAR echoes) sequence consisting of 12 periods $\Delta_k$.
The sequence was optimized for
$k_a=J$ and $k_c/k_a=0.75$ and a maximum rf amplitude of $67 \ J$.
} 
\end{figure}
\end{center}

Fig. 5 A
shows the buildup of antiphase vectors $r_2$  for 11
different offset frequencies in the range of $\pm 5 J$ (corresponding to $\pm 1$ kHz for $J \approx 200$ Hz)
during a  CROP  sequence consisting of 12 periods $\Delta_k$ without STAR echoes.
As expected, the optimal transfer efficiency is only achieved for spins close to resonance.
In contrast, 
a corresponding BB-CROP experiment with STAR refocusing produces efficient polarization transfer for a
large range of offsets (c.f. Fig. 5B).
Figure 6 shows
how the BB-CROP sequence "locks" the angle $\gamma$ between $l_2$ and
$l_1$ (c.f. Fig. 2) near its optimal value as given by on-resonance CROP pulse.

We have
carried out extensive simulations to study the loss in efficiency due to a large number of
180$^\circ$ pulses for realistic as well as hypothetical values of rf amplitudes. Fig. 7
illustrates how the offset dependence of the transfer efficiency $\eta$  is effected by
increasing the number of STAR echo periods  both in
the absence and presence  of rf inhomogeneity. From the figures it is
clear that one has to find an optimal number of evolution periods that gives the best performance
for given system parameters like maximum rf amplitude,  rf inhomogeneity and the bandwidth one
desires to cover.

\begin{center}
\begin{figure}[h] 
\centerline{\psfig{file= \fig/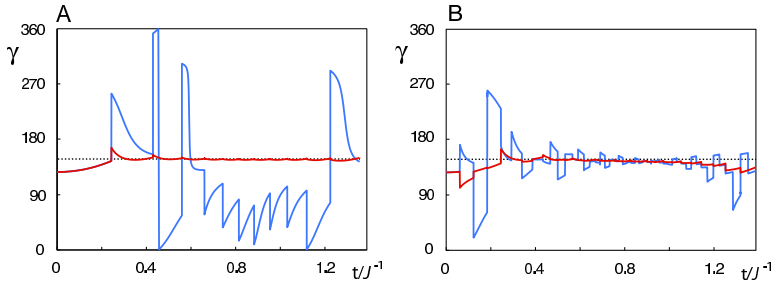 ,width=4in}}
\caption{\label{fig:beta_etc6} 
Evolution of the angle $\gamma$ during the selective CROP (A) and
BB-CROP (B) sequence as in Fig. 5 for the on-resonance case (red curves) and for
an offset of -3 $J$.
The optimal value of $\gamma$ to be maintained
during the CROP trajectory is indicated by dashed lines.
} 
\end{figure}
\end{center}

It is important to note that with high-resolution spectrometers, equipped with more rf power, 
relaxation losses during pulse periods can be made
very small. This is illustrated in Figs. 7 A and B, assuming a maximum rf amplitude on the $I$
channel of $500\ J$ and 67$J$, respectively, corresponding to 180$^\circ$ pulse durations of 5 $\mu$s and 39 $\mu$s (typical
value for $^{13}$C pulses) for the $J \approx 194$ Hz coupling constant  of the
$^{13}$C-$^{1}$H spin pair of a model system \cite{Crop, Rope}, ({\it vide infra}). For short
180$^{\circ}$ pulses (large rf amplitude) during which relaxation losses become small, a larger
number of refocusing pulses has the largest bandwidth and approaches the ideal CROP efficiency
most closely (c.f. green curve in Fig. 7 A).

The refocusing sequence  ${\Delta \over 4}  R_3 {\Delta \over 4} R_1 {\Delta \over 4} R_3
{\Delta
\over 4}$  as described in the theory section is not the only STAR refocusing
scheme for making CROP sequences broadband. For example, 
${\Delta \over 4}  R_3 {\Delta \over 4} R_2 {\Delta \over 4} R_3 {\Delta
\over 4}$
or
${\Delta \over 4}  R_2 {\Delta \over 4} R_1 {\Delta \over 4} R_2
{\Delta
\over 4}$
 will also perform STAR refocusing. However as indicated above, in practice it
may be necessary to have $\Delta$ as large as possible, in which case one should try to refocus the
largest of the components $a$, $b$, $c$ of the chemical shift generator $I_z$ (c.f. Eq.
\ref{eq:izdecomp}) more often during the refocusing cycle $\Delta$.
For example, the choice of the refocusing cycle presented in the paper is optimal for the values of
$k_c/k_a=0.75$ and
$k_a/J=1$, in which case the vector $e_3$ is mostly in the x-y plane
and hence the magnitude of component $c$ is smaller than the magnitude of $a$ or $b$. Therefore it
is of advantage to refocus $a$ and $b$ more often by performing $R_3$ rotations (the $R_3$ rotation
refocuses the $a$ and $b$ components) and hence the choice of the sequence. Since there are no
rotations of spin $S$ during the application of $R_3$ pulses, the total number of pulses and the
resulting effects due to rf inhomogeneity are minimized. Dephasing losses due to rf inhomogeneity
of the 180$^\circ$ pulses (e.g. applied to spin $S$) can be further reduced by choosing appropriate
phase cycling schemes
\cite{MLEV, XY, Fixedpoints}.

\begin{center}
\begin{figure}[h] 
\centerline{\psfig{file= \fig/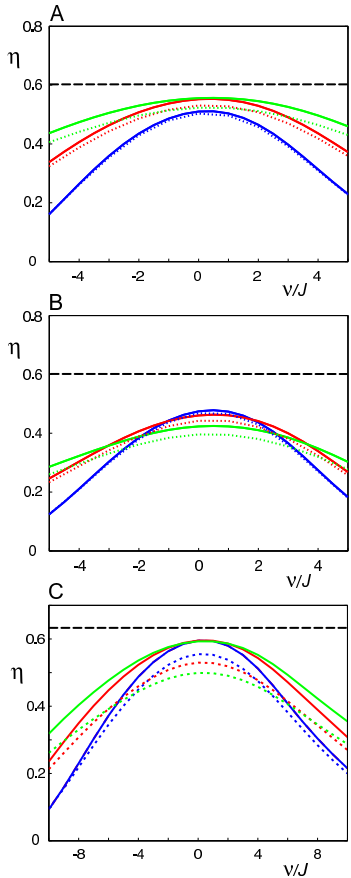 ,width=1.8in}}
\caption{\label{fig:beta_etc7} 
Offset dependence of the transfer efficiency $\eta$ for system parameters
  corresponding to the $^{13}$C-$^{1}$H moiety of 
$^{13}$C sodium formate in glycerol with $k_a=J$ and $k_c/k_a=0.75$ \cite{Crop} (A, B) and
corresponding to the $^{1}$H-$^{15}$N moiety of a protein with a 
a rotational correlation time of 70 ns with $k_a=0.8\ J$ and $k_c/k_a=0.73$   
\cite{CRINEPT} (A and B). A  maximum rf amplitude on the $I$
channel of 500
$J$ (A), $67\ J$ (B), and  550 $J$ (C) is assumed, corresponding to a  hypothetical  180$^\circ$
($^{13}$C) pulse duration of 5
$\mu$s (A), realistic on-resonance
180$^\circ$ ($^{13}$C) pulse duration of 39 $\mu$s (B) and a 180$^\circ$($^{1}$H) pulse duration
of 10 $\mu$s (C) for a $^{1}$H-$^{15}$N coupling of $J=93$ Hz. Blue, red and green curves
represent BB-CROP sequences with 4, 8 and 12 STAR echo periods $\Delta_k$, respectively (for
details, see {\it Supporting Methods}). Solid and dashed curves correspond to simulations in the
absence and presence of rf inhomogeneity, respectively, assuming a Gaussian rf distribution with a
full width at half height of 10\%. }
\end{figure}
\end{center}
In many cases it might also be possible to cut down relaxation losses by
suitable implementation of the 180$^\circ$ pulses. For
example, in the presence of a large contribution of the dipole-dipole mechanism to the
transverse relaxation rates, synchronization
of $I$ and $S$ rotations can be used to create
transverse bilinear operators such as $I_x S_x$  which commute with $I_z S_z$. This way some of
the losses might be prevented when the antiphase magnetization is passed through the transverse
plane during its inversion by $R_3$ pulses.

\vskip 4em

\section{Experimental results}

In order to test the BB-CROP pulse sequence, we chose an established model system
\cite{Crop, Rope}, consisting of  a small molecule ($^{13}$C-labeled sodium
formate) dissolved in a highly viscous solvent (($^2$H$_8$) glycerol) in order to simulate the rotational
correlation time of a large protein. Both the simplicity and sensitivity of the model system
makes it possible to quantitatively compare the transfer efficiency of pulse sequences and to
acquire detailed offset profiles in a reasonable time.
Because of its large chemical shift anisotropy and the resulting CSA-DD cross-correlation effects,
we use the 
$^{13}$C spin of $^{13}$C-sodium
formate to represent spin
$I$ and the attached
$^{1}$H spin to represent spin $S$ with a heteronuclear scalar coupling constant of  $J=193.6$ Hz.
At a temperature of 270.6 K and a magnetic
field of 17.6 T, the experimentally determined auto and cross-correlated relaxation
rates of spin $I$ were
 $k_a\approx J$ and $k_c\approx0.75 \ k_a$ (solvent: 100\% ($^2$H$_8$) glycerol).
For a given pulse sequence element, the achieved transfer efficiency of  $^{13}$C polarization $I_z$ to
$2I_z S_z$ was measured by applying 
a hard 90$_y^\circ$ proton pulse 
and recording the resulting proton
anti-phase signal  (initial $^1$H
magnetization was dephased by applying a 90$^\circ$ proton pulse followed by a pulsed magnetic
field gradient) \cite{Rope}.

\begin{center}
\begin{figure}[h] 
\centerline{\psfig{file= \fig/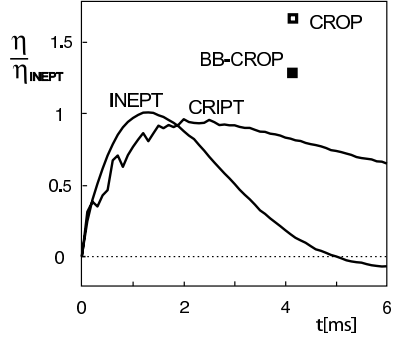 ,width=2.0in}}
\caption{\label{fig:beta_etc8} 
Experimental on-resonance transfer efficiencies of the  CROP (open square) and corresponding
BB-CROP (filled square) sequence consisting of four periods $\Delta_k$ (without and with STAR
echoes) and a total duration of 4.2 ms. For comparison, experimental on-resonance INEPT and CRIPT
transfer efficiencies are shown as a function of the transfer time. 
In
the experiments,  spins $I$ and $S$ correspond to
$^{13}$C and 
$^{1}$H in
$^{13}$C-sodium
formate dissolved in ($^2$H$_8$) glycerol. }
\end{figure}
\end{center}

Fig. 8 shows experimental on-resonance transfer efficiencies of the conventional INEPT
\cite{INEPT2} and CRIPT
\cite{cript3} sequences as a function of the mixing time. The figure also shows the 
on-resonance transfer efficiency of a CROP sequence consisting of four periods $\Delta_k$ (without refocusing)
which shows a gain of 65\% compared to the maximum INEPT efficiency. As expected (c.f.  blue curve
in Fig. 7 B), the broadband version of this sequence (BB-CROP) with four STAR echoes has a reduced
transfer efficiency because of relaxation losses during the additional 180$^\circ$ pulses, which
in the current experiments had relatively long durations due to the relatively small rf amplitude
(13 kHz) of the $I$ channel ($^{13}$C) (the BB-CROP pulse sequence is provided in {\it Supporting Methods}). Additional losses are caused by dephasing due to rf
inhomogeneity, which is typically larger for the $^{13}$C channel (where most 180$^\circ$ pulses
are given) compared to the
$^{1}$H channel.
The experimentally
determined on-resonance transfer  efficiency of BB-CROP is 28 \% larger than the maximum INEPT
transfer efficiency. In Fig. 9, the experimental offset profiles
of the $I_z \rightarrow 2 I_z S_z$ transfer efficiency of
BB-CROP and INEPT are compared. A reasonable match is found between the experiments and the
simulations shown in Fig. 7 B.

\begin{center}
\begin{figure}[h] 
\centerline{\psfig{file= \fig/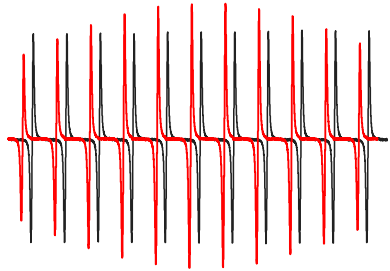 ,width=2.0in}}
\caption{\label{fig:beta_etc9} 
Experimental offset dependence of the $I_z \rightarrow 2 I_z S_z$ transfer efficiency for a
BB-CROP sequence consisting of 4 periods with STAR echoes (red) and the INEPT sequence (black).
The resulting two-spin order
$2 I_z S_z$ was converted to antiphase coherence
$2 I_z S_x$ by a hard $90^\circ (S)$ pulse and the resulting antiphase
 signals are shown for 11 offsets of spin $I$ in the range of $\pm$ 500 Hz. }
\end{figure}
\end{center}

\section{Conclusion}
In this paper we introduced the principle of specific trajectory adapted refocusing (STAR), which was used to 
design broadband relaxation optimized BB-CROP pulse sequence.
We would like to emphasize again that with increasing rf amplitudes, the efficiency of
the on-resonance cross-correlated relaxation optimized pulse can be closely approached by 
the BB-CROP sequences. As future spectrometers are
equipped with more rf power, we can significantly reduce the duration of 180$^\circ$ refocusing
pulses, which are the major bottleneck in BB-CROP achieving the maximum efficiency. Based
on our simulations, we expect
immediate gains in NMR spectroscopy of large proteins by use of the proposed BB-CROP pulses. For
example, in the HSQC experiment involving $^1$H and $^{15}$N, with maximum
rf amplitudes corresponding to 12 $\mu$s $^1$H 180$^\circ$ pulses and 40 $\mu$s $^{15}$N
180$^\circ$ pulses, we expect up to 70\% enhancement in sensitivity over a reasonable bandwidth
compared to state of the art methods.
With currently available rf amplitudes, in many applications it might even be advantageous to use
broadband versions of ROPE or optimal CRIPT (special case of CROP where $J=0$).
In these cases, we only use 180$^\circ$ pulses in the center of each evolution period and
hence loose less due to relaxation during the pulses (of course, as pointed out earlier, in these
cases in order to do a broadband
transfer, we will necessarily eliminate
either J couplings or cross-correlation ). In Figs. 8 and 9 we have not compared the sensitivity of BB-CROP with CRINEPT
\cite{CRINEPT} as the latter is not broadband for the transfer $I_z \rightarrow 2 I_z S_z$.
Similar to the on-resonance CROP pulse, we have found that the BB-CROP pulse sequence is robust to
variations in relaxation rates. Finally, the ability of the BB-CROP sequence to achieve the maximum
possible transfer efficiency over a broad frequency range by use of high rf power provides a
 strong motivation
 to build high-resolution spectrometers with short 180$^\circ$ pulses.

\newpage

\begin{center}\section{Supporting Methods}\end{center}

\subsection{Orthogonality of inphase and antiphase magnetization along CROP trajectories}

\noindent Let $r_1(t), r_2(t)$ represent the inphase and antiphase magnetization vectors as defined in the paper. Let $l_1$ and $l_2$ represent their transverse components as shown in Fig. 1 of the main text. For the sake of simplicity of notation, we will also use $l_1$ and $l_2$ to denote the magnitudes of these transverse vectors, as the true meaning will be clear from the context. In \cite{Crop}, it was shown that the CROP transfer $I_z \rightarrow 2I_zS_z$
has the following properties. Throughout the transfer, the angle $\gamma$ between vectors 
$l_1$ and $l_2$ is constant and the ratio $\frac{l_2}{l_1}$ is constant at the value $\eta$, where $\eta$ is the 
optimal efficiency of the CROP pulse. The two constants of motion completely determine the amplitude $A$ and the phase $\phi$, the CROP pulse makes with the vector $l_1$. Furthermore $\gamma$ and $\eta$ satisfy \cite{Crop} 
\begin{equation}\label{eq:etagamma}
\frac{1}{\eta}\cos(\theta - \gamma) + \eta \cos(\theta + \gamma ) = \frac{2 \xi}{\chi},
\end{equation}where $\xi = \frac{k_a}{J}$, $\chi = \sqrt{1 + \frac{k_c^2}{J^2}}$ and $\theta=\tan^{-1}(\frac{J}{-k_c})$.

Using $z_1(t) = \langle I_z \rangle (t) $ and $z_2(t) = \langle 2I_zS_z \rangle (t)$, the inner product $J$ between $r_1(t)$ and $r_2(t)$ can be expressed as $$J = z_1(t)z_2(t) + l_1(t) l_2(t)\cos(\gamma). $$ We now compute $\frac{dJ}{dt}$ along the CROP trajectory using the following identities, where the time dependence of the quantities, $ l_1, l_2, z_1, z_2, A, \phi $ is implicit.

\begin{eqnarray*}
\frac{dz_1}{dt} &=& -2 \pi A l_1\sin \phi \\  
\frac{dz_2}{dt} &=& 2 \pi A \ l_2 \sin(\gamma - \phi) \\
\frac{dl_1}{dt} &=& 2 \pi A \ z_1 \sin\phi - \pi J ( \xi\ l_1 - \chi\ l_2 \cos(\theta + \gamma)) \\
\frac{dl_2}{dt} &=& -2 \pi A \sin(\gamma - \phi)\ z_2 - \pi J ( \xi \ l_2 - \chi \ l_1 \cos(\theta - \gamma))\end{eqnarray*}

\begin{equation*} \frac{dJ}{dt} = l_1l_2 \{ A \sin\gamma(\frac{z_1}{l_1}\cos\phi -\frac{z_2}{l_2}\cos(\gamma - \phi))   
+ J \chi\ ( \frac{l_1}{l_2} \cos(\theta - \gamma) + \frac{l_2}{l_1} \cos(\theta + \gamma ) - 2 \frac{\xi}{\chi}) \}.
\end{equation*}

\noindent Along the CROP trajectory, $\frac{d \gamma}{dt} = 0$ implies \cite{Crop} $$ A \sin\gamma\ (\frac{z_1}{l_1}\cos\phi -\frac{z_2}{l_2}\cos(\gamma - \phi)) = 0. $$ Also along the CROP trajectory $\frac{l_2}{l_1} = \eta$. Using (Eq. \ref{eq:etagamma}), we then obtain that along CROP trajectory 
$$ \frac{l_1}{l_2} \cos(\theta - \gamma) + \frac{l_2}{l_1} \cos(\theta + \gamma ) = \frac{2 \xi}{\chi}. $$
This then implies $\frac{dJ(t)}{dt} = 0$. Since $J(0) = 0$ as $r_2(0) = 0$, we obtain that throughout the transfer
$J(t) = 0$.

\subsection{Dynamic Programming method for finding optimal sequence of flips and
delays}

We now explain the method of dynamic programming \cite{Bellman} for finding the optimal sequence
of pulses and delays that best approximates relaxation optimized pulse sequences. The method is
best illustrated by considering  the simpler case when there is no cross-correlation in the
system. In absence of cross-correlation, the relaxation optimized transfer of $I_z \rightarrow
2I_zS_z$ is characterized by gradual rotation of the operator $I_z \rightarrow I_x$, followed by
the rotation  $2I_yS_z \rightarrow 2I_zS_z$ \cite{Rope}.  Let $r_1$ be the magnitude of in-phase
terms, i.e.,
$r_1^2 = \langle I_x\rangle^2 + \langle I_z\rangle^2$. Let $\beta_1$ be the angle $r_1$ makes with the transverse plane, i.e. $\beta_1=\cos^{-1}{{\langle I_x \rangle}\over{r_1}}$
(see Fig. 1). 
Let $r_2$ measure the magnitude
of the total antiphase terms, i.e.,  $r_2^2 = \langle
2I_yS_z\rangle^2 + \langle 2I_zS_z\rangle^2$ and let $\beta_2 = \cos^{-1}{{\langle 2 I_y S_z \rangle}\over{r_2}}$
(see Fig. 1).
Using rf fields, we can exactly control  the angle
$\beta_1$ and $\beta_2$ and these are thought of as control parameters (see Fig. 1). During the evolution of relaxation optimized pulse sequence \cite{Rope} one of the $\beta_1$ or $\beta_2$ is zero, so we assume $(\beta_1, \beta_2) \in ([0, \frac{\pi}{2}], 0) \cup (0, [0, \frac{\pi}{2}])$.

\begin{center}
\begin{figure}[h]
\centerline{\psfig{file= 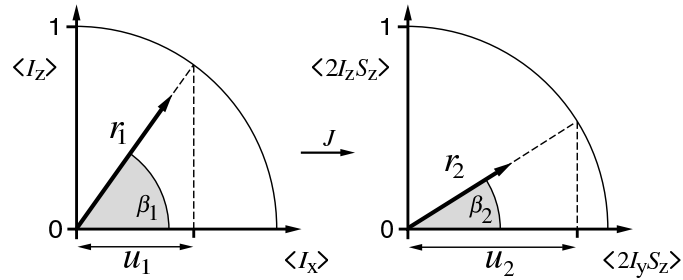 ,width=3.5 in}}
\caption{ Representation of the system variables $r_1$, $r_2$, the angles
$\beta_1$,
$\beta_2$, and of the control parameters
$u_1=\cos \beta_1$, $u_2=\cos \beta_2$ in terms of the expectation values $\langle I_x \rangle$, $\langle I_z
\rangle$, $\langle 2 I_y  S_z \rangle$, and $\langle 2 I_z  S_z \rangle$.}
\end{figure} 
\end{center}

Now suppose, we only have one evolution period,  consisting of a pulse and delay, at our disposal. We can compute this optimal pulse and delay so that at the end of the evolution period, $r_2$ is 
maximized. Starting with $r_1, r_2$ as defined and a given choice of $\beta_1$, $\beta_2$ and $\tau$, the values of $r_1(\tau)$ and $r_2(\tau)$ at 
the end of the period $\tau$ are \begin{eqnarray} \label{eq:r1} r_1^2(\tau) &=& \exp(- 2 \pi k_a \tau) (r_1 \cos\beta_1\cos(\pi J \tau) - r_2\cos\beta_2\sin(\pi J \tau))^2 + (r_1 \sin\beta_1)^2 \\
\label{eq:r2} r_2^2(\tau) &=& \exp(- 2 \pi k_a \tau) (r_2 \cos\beta_2\cos(\pi J \tau) + r_1\cos\beta_1\sin(\pi J \tau))^2 + (r_2 \sin\beta_2)^2. \end{eqnarray} We write this as $(r_1(\tau), r_2(\tau)) = f(r_1, r_2; \beta_1, \beta_2, \tau)$. We can maximize the expression in equation (\ref{eq:r2}) and find the optimal value of $\tau, \beta_1, \beta_2$ and also the largest achievable value  $r_2(\tau)$. This value depends only on the initial value $r_1$ and $r_2$ and we call it $V_1(r_1, r_2)$, the optimal return function at stage 1 starting from $r_1, r_2$. This optimal return function represents the best we can do starting from a given value of $r_1$ and $r_2$ given only one evolution period. 

Given two evolution periods, then by definition $V_2(r_1, r_2) = \max_{\beta_1, \beta_2, \tau}
V_1(f(r_1, r_2; \beta_1, \beta_2, \tau))$. The basic idea is, since we have computed $V_1$ for various values of $r_1$ and $r_2$ , we can use it to compute $V_2$. In general then

\begin{equation} \label{eq:Dynamic}
V_n(r_1, r_2) = \max_{\beta_1, \beta_2, \tau} V_{n-1}(f(r_1, r_2; \beta_1, \beta_2, \tau)).
\end{equation}

\noindent Thus the dynamic programming proceeds backwards. We first compute the optimal return functions $V_1$ followed by $V_2$ and so on. Computing $V_k(r_1, r_2)$, also involves computing the best value of the control parameters $\beta_1, \beta_2, \tau$ to choose for a given value of state $r_1, r_2$ at stage $k$. We denote this optimal choice as $\beta_1^\ast(r_1, r_2, k), \beta_2^\ast(r_1, r_2, k), \tau^\ast(r_1, r_2, k)$, indicating that the optimal control depends on the state $r_1, r_2$ and the stage $k$.  

\noindent In practice, the algorithm is implemented by sampling the $[0,1] \times [0,1]$ square in the $r_1, r_2$ plane uniformly into say 100 points. Each of these points correspond to a different value of $r_1, r_2$. By maximizing the expression in equation $\ref{eq:r2}$, we can compute the optimal $\beta_1, \beta_2, \tau$ for each of these points. This would give us $\beta_1^\ast(r_1, r_2, 1), \beta_2^\ast(r_1, r_2, 1), \tau^\ast(r_1, r_2, 1)$ and also $V_1(r_1, r_2)$. Now to find $V_2(r_1, r_2)$ at these points, we sample the control space $(\beta_1, \beta_2) \in ([0, \frac{\pi}{2}], 0)\cup(0, [0, \frac{\pi}{2}])$ and $\tau \in [0, \frac{1}{2J}]$ uniformly and compute the value $f(r_1, r_2, \beta_1, \beta_2, \tau)$ for all of these samples $\beta_1, \beta_2, \tau$ and choose the one that has the largest value $V_1(f(r_1, r_2; \beta_1, \beta_2, \tau))$ .This then is the best choice of control parameters if there are two evolution periods to go. We also then obtain $V_2(r_1, r_2) = V_1(f(r_1, r_2, \beta_1^\ast, \beta_2^\ast, \tau^\ast))$. We can continue this way and compute $V_n(r_1, r_2)$. Now to construct the optimal pulse sequence consisting of $N$ evolution periods, we just look at the value $\beta_1^\ast(1, 0, N), \beta_2^\ast(1, 0, N), \tau^\ast(1, 0, N)$ and evolve the system according to these parameters and get $r_1$ and $r_2$ at beginning of stage $N-1$. But we also know $\beta_1^\ast(r_1, r_2, N-1), \beta_2^\ast(r_1, r_2, N-1), \tau^\ast(r_1, r_2, N-1)$ which is then used to evolve the system for one more step and so on. From the sequence $\beta_1^\ast(r_1, r_2, k), \beta_2^\ast(r_1, r_2, k), \tau^\ast(r_1, r_2, k)$, \ $k=1...N$, the optimal flip angles can be immediately determined.

\newpage

\subsection{BB-CROP Pulse sequence parameters}

\noindent
{\bf Table 1:} Parameters of a BB-CROP sequence (c.f. Fig. 4 A and D)
consisting of four STAR
echoes optimized for
$k_a/J=1$, $k_c/k_a=0.75$ and amplitude $A=67 \ J \approx 13$ kHz  of the $I$
channel for
$J=193.6$ Hz. The DANTE-type on-resonance pulses applied to spin $I$ are
denoted $\alpha_k$. The
Table only specifies explicitly the parameters for the spin $I$ rf channel, however
note that the pulse
elements $R_1(t)$ require a simultaneous hard 180$^\circ$ rotation of spin $S$
around an axis in
the transverse plane. In our experiments, the phases of the four 180$^\circ(S)$
pulses were chosen
according to the XY-4 cycle 0$^\circ$, 90$^\circ$, 0$^\circ$, 90$^\circ$
\cite{XY} in order to
reduce the effects of rf inhomogeneity of the $S$ pulses. 
\smallskip
\smallskip

\vskip 1cm

\center
\begin{tabular}{c c c c c }

type & duration [$\mu$s] & offset [kHz] & Phase [deg] 
\\  \hline 
$\alpha_0$      &   5.9   &   -   &    0  \\ \hline
$\Delta_1/4$&   300.2   &  -    &    -  \\
$R_3(t)$           &   36.1   & $-4.77$     &   119.5   \\
$\Delta_1/4$&   300.2   &  -    &    -  \\
$R_1(t)$           &   12.7   &   $37.13$   &   20.9   \\
$\Delta_1/4$&   300.2   &  -    &    -  \\
$R_3(t)$           &   37.6   &   $-2.76$   &   224.6   \\
$\Delta_1/4$&   300.2   &  -    &    -  \\ \hline

$\alpha_1$      &  7.9    &  -    &    393.9  \\ \hline

$\Delta_2/4$&   220.2   &  -    &    -  \\
$R_3(t)$       &34.1        &   -6.82   &    182.5  \\
$\Delta_2/4$&   220.2   &  -    &    -  \\
$R_1(t)$      &25.1          &  15.13    &    36.3  \\
$\Delta_2/4$&   220.2   &  -    &    -  \\
$R_3(t)$     & 35.5         &   -5.39   &    223.5  \\
$\Delta_2/4$&   220.2   &  -    &    -  \\ \hline

$\alpha_2$      &   8.6   &  -    &  340.5    \\ \hline

$\Delta_3/4$&   219.7   &  -    &    -  \\
$R_3(t)$     &34.5          &   -6.42   &   157.3   \\
$\Delta_3/4$&   219.7   &  -    &    -  \\
$R_1(t)$      &35.2          &   5.74   &   353.8   \\
$\Delta_3/4$&   219.7   &  -    &    -  \\
$R_3(t)$       &34.4        &    -6.48  &   136.6   \\
$\Delta_3/4$&   219.7   &  -    &    -  \\ \hline

$\alpha_3$      &   7.9   &  -    &  217.3    \\ \hline
$\Delta_4/4$&   301.2   &  -    &    -  \\
$R_3(t)$       &36.7        &  -4.04    &  58.3    \\
$\Delta_4/4$&   301.2   &  -    &    -  \\
$R_1(t)$      & 38.3          &   -1.19   &  269.0    \\
$\Delta_4/4$&   301.2   &  -    &    -  \\
$R_3(t)$       &36.4         &  -4.49    &  344.1    \\
$\Delta_4/4$&   301.2   &  -    &    -  \\ \hline

$\alpha_4$      &  4.2    &  -    &  350.3    \\ \hline

\end{tabular}
\end{document}